\documentclass[%
 reprint,
superscriptaddress,
 amsmath,amssymb,
 aip,
floatfix,
]{revtex4-1}

\usepackage{graphicx}
\usepackage{xcolor}
\usepackage[english]{babel}
\usepackage{dcolumn}
\usepackage{bm}
\usepackage{bm}
\usepackage[utf8]{inputenc}
\usepackage[T1]{fontenc}

\newcommand{\TDEP}{\textsc{TDEP}}
\newcommand{\diff}{\mathrm{d}}

\newcommand{\kdotp}{ {\mathbf{k} \cdot \mathbf{p}} } 
\newcommand{\sintef}{SINTEF Industry, NO--0314 Oslo, Norway}
\newcommand{\OsloC}{Centre for Materials Science and Nanotechnology,  Department of Physics,  University of Oslo,  NO-0316 Oslo, Norway}
\newcommand{\NMBU}{Department of Mechanical Engineering and Technology
Management, Norwegian University of
Life Sciences, NO-1432 \AA s,
  Norway.}

\newcommand{\VASP}{\textsc{VASP} }

\begin{document}


\title{Discarded gems:
Thermoelectric performance of materials with band gap emerging at the hybrid-functional
level}


\author{Kristian Berland}
\email{kristian.berland@nmbu.no}
\affiliation{\NMBU}
\author{Ole Martin L\o vvik}
\affiliation{\sintef}
\affiliation{\OsloC}
\author{Rasmus Tran\aa s}
\affiliation{\NMBU}

\date{\today}

\begin{abstract}
A finite electronic band gap
is a standard filter in high-throughput screening of materials using density functional theory (DFT).
However, because of the systematic underestimation of band gaps in standard DFT
approximations, a number of compounds may incorrectly be predicted metallic.
In a more accurate treatment, such materials may instead appear as low band gap
materials and could e.g. have good thermoelectric properties if suitable doping is feasible.
To explore this possibility, we performed hybrid functional calculations on
1093 cubic materials
 listed in the \textsc{MaterialsProjects} database with four atoms in the primitive unit
cell, spin-neutral ground state, and a formation energy within 0.3 eV of the convex hull.
Out of these materials, we identified eight compounds for which a finite band gap emerges.
Evaluating electronic
and thermal transport properties of these compounds, we found the compositions
MgSc$_2$Hg and Li$_2$CaSi to exhibit promising thermoelectric properties.
These findings underline the potential of reassessing band gaps and band structures
of compounds to indentify additional potential thermoelectric materials.
\end{abstract}

\maketitle


Thermoelectrics, with their ability to turn temperature gradients into electricity,
can contribute to making
the transition into a green economy with
reduced greenhouse emission by
recovering some of the waste heat generated in various
industrial processes.\cite{Rowe1998_review,Dresselhaus2013_review,Mahan2016_review}
While thermoelectric materials have traditionally not been sufficiently
efficient for this task, great strides forward have been made in recent
years. This has in turn intensified the hunt for novel thermoelectric
materials,\cite{bhattacharya_high-throughput_2015,bhattacharya_novel_2016,chen_understanding_2016-1,li_high-throughput_2019,
barreteau_looking_2019,chen_machine_2019-1, iwasaki_machine-learning_2019}
including the adoption of high-throughput screening  and material
informatics \cite{butler_machine_2018,hillMaterialsScienceLargescale2016}
approaches.

The thermoelectric figure-of-merit $ZT= \sigma S^2 T/(\kappa_e + \kappa_\ell)$,
which is measure of the conversion efficacy,
is given by the conductivity $\sigma$, the Seebeck coefficient $S$,
the electronic $\kappa_e$, and lattice thermal $\kappa_\ell$ conductivity.
Among these, all but $\kappa_\ell$ are strongly linked to the electronic band
structure.
The electronic band gap $E_{\rm gap}$ is a particularly important parameter,
determining the temperature for the onset of minority carrier
transport, which causes a marked drop in $S$. It also has an indirect influence
on the band curvature, i.e. as revealed by
$\mathbf{k}\cdot\mathbf{p}$-theory.\cite{luttinger_quantum_1956}
Following Sofo and Mahan,\cite{sofo_optimum_1994-1} a band gap of approximately 6-10 $k_{\rm B}T$ has traditionally been considered
attractive.
However, their analysis was based on a direct band-gap model with a single
valley. Given its link to the band curvature, the band gap $E_{\rm gap}$ can also be viewed as a scale factor making a low band
gap material more prone to exhibit multiple valleys in multipocketed band structures;\cite{wang_band_2021}
nonetheless, the need to limit bipolar transport has made the existence of
a finite band gap a standard criterion in most screening
studies.\cite{li_high-throughput_2019}
Recently, attention has been broadened to other types of materials:
Semi-metals with a strong  asymmetry between conduction and valence
bands have e.g. been marked as potential thermoelectric
materials.\cite{markov_semi-metals_2018,markov_thermoelectric_2019}
Gapped metallic systems, which possess a band gap within the conduction or
valence band, could also
potentially exhibit good thermoelectric properties,
once the band edge is sufficiently doped towards the Fermi level.\cite{ricci_gapped_2020}

A completely different reason for not discarding predicted metallic systems is that a
number of them might have been mislabeled due to various approximations used in
density functional theory (DFT).\cite{malyi_false_2020}
In particular, the commonly used generalized gradient approximation (GGA) systematically
underestimates band gaps.\cite{pribram-jones_dft_2015,morales-garcia_empirical_2017} This  is less the case for hybrid
functionals,\cite{kim_band-gap_2020} which mix a fraction of "exact" Fock exchange with the GGA.\cite{HSE03,HSE06}
In the empirical linear relations
between experimental and computed band gaps  of Morales-García et
al.,\cite{morales-garcia_empirical_2017}
the offset of about
 0.92 eV roughly  indicates that compounds with a band gap smaller than this
are likely to be incorrectly predicted as metallic by GGA.

In this work, we computed the band gap of 1093 cubic nonmagnetic materials
listed in the \textsc{Materials Project} database
\cite{jain_commentary_2013} with
four atoms in the primitive unit cell and a formation energy within 0.3 eV of
the convex hull.
These compounds include the full Heusler compounds with
spacegroup $Fm{\bar{3}}m$, inverse Heuslers with spacegroup $F{\bar{4}}3m$ (both
with composition  $X_2YZ$) and binary $AB_3$ compounds.
This reassessment resulted in eight compounds that were possibly mislabeled
metallic by GGA.
DFT calculations were performed using the
\VASP\cite{vasp1,vasp3,vasp4,gajdos_linear_2006-1} software package.
The consistent-exchange van der Waals functional vdW-DF-cx
functional \cite{berland_exchange_2014,berland_van_2015} was used
for obtaining relaxed crystal structures and lattice thermal conductivities. While mostly used for modelling non-covalently bonded
solids, recent studies have shown that vdW-DF-cx can improve structure and
energetics compared to that of GGA of ionic and covalently-bonded structures as well.\cite{lindroth_thermal_2016,gharaee_finite-temperature_2017,hyldgaard_screening_2020}
To identify materials that could possess a band gap at the hybrid level, we
first computed the band gap using merely a $4\times 4\times 4$
$\mathbf{k}$-sampling of the Brillouin zone including spin-orbit coupling
using the HSE06 \cite{HSE03, krukau_influence_2006} hybrid functional.
Such a low sampling can result in inaccurate Kohn-Sham energies and we
acknowledge that there is a
slight risk that some compounds with very low band gap are missed.
But generally, the coarse sampling will cause a few systems to incorrectly
appear with a finite or too large
band gap. All systems with a finite band gap in the first stage were therefore reassessed
with a $12\times 12\times 12$ $\mathbf{k}$- Brillouin-zone sampling of the Fock operator and charge density, which is used to compute the band structure path using 101 $\mathbf{k}$-points along $W$-$L$-$\Gamma$-$X$-$K$ to obtain an accurate band gap.
For the new band-gap compounds, the electronic transport properties were computed with the Boltzmann Transport
equation in the constant relaxation time approximation with $\tau = 10^{-14}{\rm
s}$ using \textsc{BoltzTraP}.\cite{madsen_boltztrap._2006}
To ensure dense grid sampling, we used a corrected $\kdotp$-based interpolation
method,\cite{berland_enabling_2017, berland_thermoelectric_2018}
using the same computational parameters as in
Ref.~\onlinecite{berland_thermoelectric_2019}
The lattice thermal conductivity, $\kappa_\ell$, was computed using the temperature-dependent effective
potential (\TDEP) method.\cite{hellman_lattice_2011-2,hellman_phonon_2014} A canonical ensemble was used to generate 50 uncorrelated configurations based on a $3 \times 3 \times 3$ repetition of the relaxed
primitive cell.\cite{shulumba_lattice_2017}
The positions and forces of the supercells allowed for extraction of second- and third-order force constants. The cutoff for the second-order interactions was set to 7 Å, while to third-order, a cutoff slightly larger than half the width of the supercell was used. Reciprocal space discretization for Brillouin zone integrations was done using a $35\times35\times35$ $\mathbf{q}$-point grid.
Isotope scattering was also included.
All supplementary GGA calculations in this letter were based on the version of Perdew-Burke-Ernzerhof (PBE).\cite{pbe1996}

\begin{ruledtabular}
\begin{table}[]
\centering
\begin{tabular}{l|ccc}
compound & \#valence & $E_{\rm hull}$ (MP) & band gap (eV) \\
\hline
 AlVFe$_2$     &24  & 0  &  0.78 \\
 Ba$_2$HgPb   &  20   & 0 &  0.06 \\
 HfSnRu$_2$  & 24 & 0 & 0.21\\
 Li$_2$CaSi & 10  & 0 & 0.01 \\
  MgSc$_2$Hg & 20  & 0 & 0.23\\
  TaInRu$_2$ & 24  & 0 & 0.05 \\
  TiSiOs$_2$ & 24 & 0 &   0.55 \\
VGaFe$_2$   & 24  & 0  & 0.66
    \end{tabular}
    \caption{Properties of new band-gap compounds}
    \label{tab:properites}
\end{table}
\end{ruledtabular}
Among the 1093 compounds materials examined, eight compounds have a band gap at the HSE06 level as listed in Tab.~\ref{tab:properites}, corresponding band structures are provided in supplementary material (SM).

\begin{figure}[t]
\includegraphics[width=0.37\textwidth]{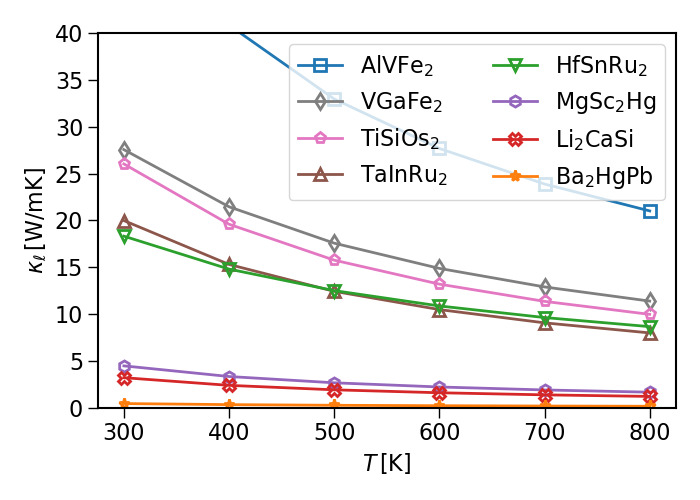}
	\caption{\label{fig:kappa} Lattice thermal conductivity of identified compounds
	computed with \TDEP.}
\end{figure}

Figure \ref{fig:kappa} shows the computed $\kappa_\ell$ for the identified compounds.
Very low values of $\kappa_\ell$ was found for  Ba$_2$HgPb
ranging from 0.46~W/mK at 300~ K to 0.17 W/mK at 800 K.
This compound was also
studied by  He et al. \cite{he_ultralow_2016} predicting values of $\kappa_\ell$
somewhat larger than ours.
Possible reasons for this differene inlude their use of a compressive sensing lattice
dynamic technique\cite{compsens_Zhou2014} to obtain third-order force constants and other technical detials,
differing exchange correlation functionals, and the phonon-mode renormalization
 inherit to \textsc{TDEP}.
Comparing \textsc{TDEP} and  \textsc{Phonopy},
Feng~et~al. \cite{feng_characterization_2020} found lower $\kappa_\ell$ for
\textsc{TDEP} than with the standard-finite difference approach and argued that \textsc{TDEP} is better suited to describe low-$\kappa_\ell$ materials.

Based solely on Fig.~\ref{fig:kappa}, only Ba$_2$HgPb, Li$_2$CaSi, and
MgSc$_2$Hg have low enough $\kappa_\ell$ to conceivably be good thermoelectric
materials.
Yet, the literature is riddled with examples of
how various disorder-related scattering mechanisms such as grain
boundaries, defects, and substitutions can dramatically lower
$\kappa_\ell$.\cite{Abdeles1963,Tian2012, Ankita:ZrNiSn,arrigoni_first-principles_2018,
carrete_nanograined_2014,Li_PRB12,SimenPaper,schrade_role_2017}
For this reason, we used $\kappa_\ell = 4\, {\rm W/mK}$ as the maximum for all
materials in further
comparisons.

\begin{figure}[t!]
\includegraphics[width=0.4\textwidth]{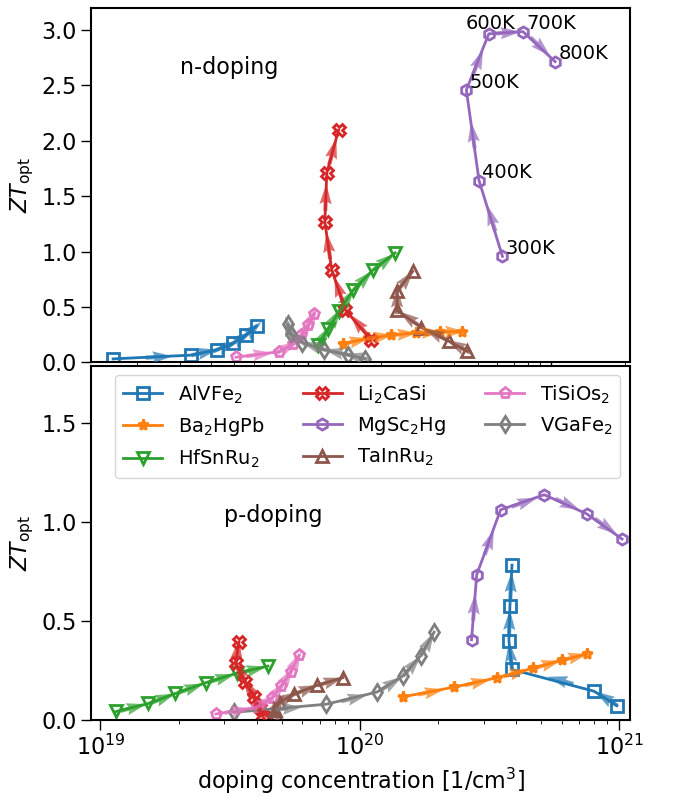}
	\caption{\label{fig:ZTopt} Optimized $ZT$ at different temperatures from
	300 K to 800 K with arrows indicating increasing temperature in steps of
	100 K. The vertical axis gives the optimized $ZT$ while the horizontal
	gives the corresponding doping concentration.}
\end{figure}

Figure~\ref{fig:ZTopt} plot the optimal doping concentration against
peak $ZT$ for each of the compounds in temperature steps of 100~K from 300~K
to 800~K, for doping concentration between $10^{18}{\rm cm^{-3}}$ and $3 \times 10^{21}{\rm
cm^{-3}}$.
Based on this plot, we deem Li$_2$CaSi and MgSc$_2$Hg to have great potential as
thermoelectric $n$-type materials, while MgSc$_2$Hg and
AlVFe$_2$ have some potential as $p$-type thermoelectrics. 
$n$-type AlVFe$_2$ has been studied earlier theoretically
at the hybrid functional level,\cite{bilc_low-dimensional_2015}
and experimentally.\cite{mikami_thermoelectric_2012,vasundhara_electronic_2008}
The study of Mikami et al. \cite{mikami_thermoelectric_2012} measured $ZT$ in a similar range
as us once doping and sublattice disorder were introduced.
While Li$_2$CaSi is reported as stable in the $Fm\bar{3}m$ Heusler phase in the
\textsc{Materials
Project}; experimentally, it has been crystallized in the orthorhombic $Pmmm$ phase.\cite{stoiber_lithium_2017}
The related Li$_2$CaSn, on the other hand, does crystallize in the Heusler phase.
No experimental realizations of MgSc$_2$Hg are known to us.

\begin{figure}[h!]
    \includegraphics[width=0.4\textwidth]{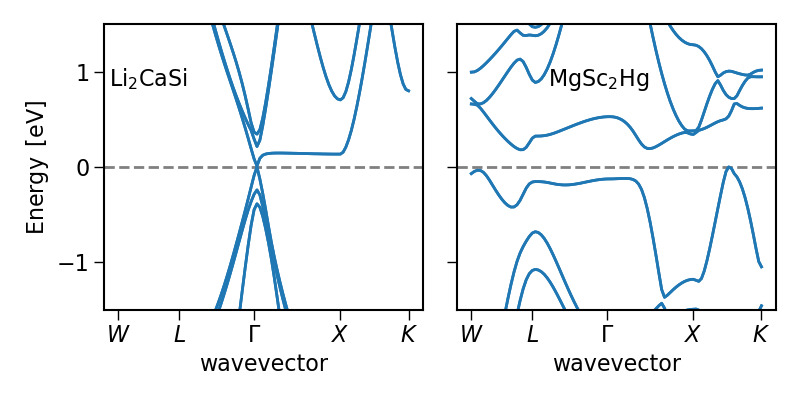}
\caption{\label{fig:bs} Electronic band structures of Li$_2$CaSi and MgSc$_2$Hg.}
\end{figure}

The origin of the high $ZT$ of Li$_2$CaSi and MgSc$_2$Hg can be related to their
band structures as shown in Fig.~\ref{fig:bs}.
The band structure of Li$_2$CaSi exhibits some noticeable features:
{\it i.} Dirac points at the $\Gamma$-point with a band opening of 0.01 eV,
{\it ii.}
near convergence of a number of additional bands at the $\Gamma$ point,
{\it iii.} electron bands that are flat in  the $\Gamma$-$X$ direction, but dispersive in the
$X$-$K$ direction. In our study, we find similar features in the band structure of HfSnRu$_2$ and
TaInRu$_2$, which also exhibit relatively high $ZT$ for $n$-type doping.
While Bilc et
al.\cite{bilc_low-dimensional_2015} argued that band structures of this type can give rise
to high $ZT$ due to the their effectively low-dimensional transport,
Park et al.\cite{park_high_2019} demonstrated that flat-and-dispersive band structures,
specifically for the case of Fe$_2$TiSi, can cause large effective scattering phase-space which significantly reduces the power factor.
In contrast, MgSc$_2$Hg band structure has a multi-valley structure in
particular in the conduction band. In fact, with the exception of the highly dispersive
band in the $X$-point, the band structure can be
viewed as a partial realization of $\delta$-function like transport spectral
function, which in the analysis of Mahan and Sofo is optimal for thermoelectric performance.\cite{mahan_best_1996}
Other cubic structures, such as the 10-valence electron full-Heusler compounds
predicted by He et al. \cite{he_designing_2019} also have similarly attractive band structure features.

\begin{figure*}[t!]

\includegraphics[width=0.80\textwidth]{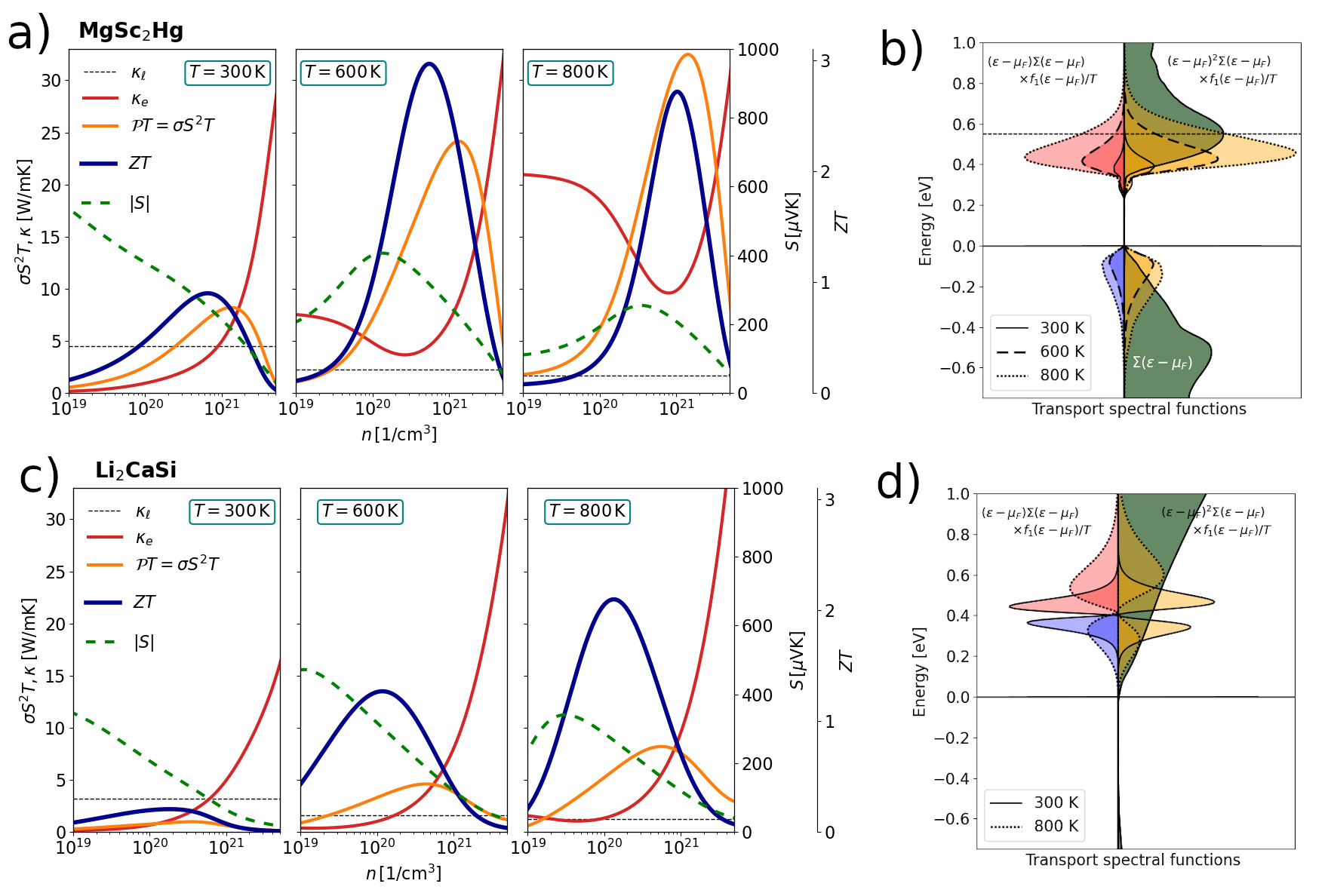}
    \caption{\label{fig:analysis} a) [c)] Thermoelectric properties of MgSc$_2$Hg [Li$_2$CaSi]
    as a function of doping doping concentration at 300 K, 600 K, and 800~K.
    In b) [d)], the green background shows the corresponding transport spectral functions $\Sigma(\epsilon)$.  The left and right side of the vertical axis show the spectral contributions to the first and second moment of the  $\Sigma(\epsilon)$ weighted by the derivative $f_1(\epsilon) = - {\rm d} f_{\rm FD} (\epsilon) {\rm d} \epsilon$, which is proportional to respectively $T \sigma S$ and the closed circuit thermal conductivity $\kappa_0$. Results for 600~K omitted for clarity in d) }
\end{figure*}

Figure~\ref{fig:analysis}a shows Pisarenko-type plots for the thermoelectric properties of  MgSc$_2$Hg at 300, 600, and 800~K.
while \ref{fig:analysis}b  shows the underlying spectral quantities giving rise to these properties.
They are related to through the transport spectral function $\Sigma(\epsilon)$ as follows\cite{madsen_boltztrap._2006}
\begin{align}
\sigma &= e^2 \int \diff \epsilon \, \Sigma(\epsilon -\mu_F) f_1 (\epsilon -\mu_F) \\
\sigma S &= (e/T) \int \diff \epsilon \, (\epsilon -  \mu_{\rm F}) \Sigma(\epsilon-\mu_F) f_1 (\epsilon-\mu_F) \\
\kappa_0 &=  (1/T) \int \diff \epsilon \, \Sigma(\epsilon-\mu_F) (\epsilon -  \mu_{\rm F})^2 f_1 (\epsilon-\mu_F), \label{eq:kappa0}
\end{align}
where $\mu_{\rm F}$ is the Fermi level and $f_1$ is the Fermi window, given by the derivative of the Fermi-Dirac function, $f_1(\epsilon - \mu_{\rm F}) = - \diff f_{\rm FD} /\diff \epsilon$. The open-circuit electronic thermal
conductivity $\kappa_{\rm 0}$ is related to the closed-circuit by $\kappa_{\rm e} = \kappa_0 - T \sigma S^2$.
The temperature dependence stems explicitly from the Fermi-Dirac function and implicitly from the temperature dependence of $\mu_F$.
A dashed line indicates the peak of $\Sigma(\epsilon)$
for comparison with the band structure in Fig.~\ref{fig:bs}.
The figures shows that for ${\rm MgSc_2Hg}$ the magnitude of $\kappa_{\rm e}$ is a key factor limiting $ZT$ at elevated temperatures.
They also show that a minimum in $\kappa_e$
at 600 and 800 K occurs at a higher doping concentration than what
maximises $S$. They both reach extreme values due to a minimum in the bipolar transport,  but the second moment $(\epsilon - \mu_F)^2$
entering into $\kappa_0$ (Eq.~\ref{eq:kappa0})  shifts the optimum of $\kappa_e$ to a higher doping concentration.
The figure also indicates that the rapidly rising $\Sigma(\epsilon)$ up to the peak occurring at 0.55
eV explains why $S$ can be quite large
despite a low band gap even at high doping concentrations.
At the same time, it shows that this rapid rise is the cause of the large values of $\kappa_e$ at high temperatures.

Figure~\ref{fig:analysis}c and  \ref{fig:analysis}d shows corresponding results for Li$_2$CaSi.
It is interesting to note that while the band gap is tiny,
the low $\Sigma(\epsilon)$ in the valence band makes this compound resemble a wide band-gap semiconductors.
In fact, at optimal doping concentration, the bipolar transport occurs almost entirely within the condution band.
While the limited bipolar transport results in higher $S$ at lower doping concentrations, Li$_2$CaSi lacks the beneficial peak in $\Sigma(\epsilon)$ present in MgSc$_2$Hg
which limits  $\kappa_{\rm e}$ at higher temperatures and doping concentrations.
The low band gap of Li$_2$CaSi makes it interesting to also consider the properties of Li$_2$CaSi as predicted at the GGA level.
In this case, a finite gap is retained at the $\Gamma$ -point but the material is self-doped and the flat-and-dispersive band crosses the Fermi level
at zero extrinsic doping. An optimal $ZT=0.76$ at 800~K is predicted -- further details in SM.


While we in this study assessed the properties of 1093 four-atom materials using
sub-converged hybrid functional calculations,
other approaches could also be worth exploring.
We investigated the potential of analyzing the GGA-level density of states,
in which a "narrowing" could hint of a finite band gap.
Details can be found in the SM.
Interestingly, this approach clearly indicated all compounds except the
MgSc$_2$Hg compound; precisely the property that made this material into a promising
thermoelectric, i.e. the high density of states close to the band
edges at the hybrid level, made the density-of-states narrowing at the GGA-level
vanish. We therefore do not generally recommend this approach to uncover high performance
thermoelectric materials.


In this letter, we have demonstrated that the use of GGA-level
band structures can cause promising thermoelectric materials to be discarded
because they are falsely predicted to be metallic.
This was illustrated with the finding of new thermoelectric compounds
with a band gap appearing at the hybrid functional level:
Out of the 1093 studied compounds, 8 were identified with a band gap by hybrid calculations and not by GGA calculations.
Out of these, a few were also promising for thermoelectric applications:
MgSc$_2$Hg, Li$_2$CaSi, and to some
extent AlVFe$_2$.
The Heusler MgSc$_2$Hg compound, in particular, exhibits excellent potential as a
thermoelectric material.
We are not aware of any experimental realization of this compound or
in-depth stability analysis. Moreover, the toxicity of Hg reduces the
attractiveness of this compound for general-purpose applications.
In addition to realizability, we stress the use of a constant
relaxation-time approximation is a coarse approximation.
The inclusion of proper electron-phonon scattering can have a
decisive impact upon the power factor and prediced $ZT$ properties.\cite{zhou_large_2018} 
Another concern is whether hybrid functionals in fact do provide
accurate band structures for these intermetallic compounds, which can be investigated for instance by performing $GW$-level calculations, as earlier done for selected Half Heuslers.\cite{zahedifar_band_2018}
Despite these caveats, our study clearly underlines that high performing thermoelectric materials
can be uncovered through reassessment of electronic band gaps.

On a final note, it is interesting that the three compounds with lowest
$\kappa_\ell$ and two of the compounds with the highest $n$-type $ZT$ violated
the octet rule or the corresponding 18- and 24-electron rules. This violation is a
feature shared with the well-known thermoelectric PbTe and related compounds. \cite{PbTe:nanostrucuring,he_designing_2019}
The existence of lone $s$-pairs have earlier been
linked to low thermal conductivity.\cite{delaire_giant_2011, jana_origin_2016,
he_designing_2019}
One could speculate that going beyond GGA could be particularly
pertinent for the electronic band structure of octet violating systems, similar to what we found earlier for
PbTe.\cite{berland_thermoelectric_2018}

\section*{Supplementary material}
See supplementary material for computed band structures at the HSE06 level, density of states at the PBE level of theory.
Band structure and n-type thermoelectric properties of Li$_2$CaSi

\section*{Acknowledgement}

The computations were performed on resources provided by
UNINETT Sigma2 - the National Infrastructure for High Performance Computing and
Data Storage in Norway. This work is in part funded by the Allotherm project (Project no. 314778) supported by the
Research Council of Norway.
Additional data beyond what is contained the article and SM
are available from the corresponding author upon reasonable request.

\bibliography{Library,extra,Libarary2,HalfHesuler_screening, Library_Rasmus,library_olem}
\end{document}